\documentclass{npw}
\usepackage{graphicx}
\usepackage{color}	

\begin{document}

\title{The suggested presence of the tetrahedral-symmetry in~the~ground-state~configuration \boldmath
of~the~$^{96}_{40}$Zr$_{56}^{}$ nucleus}
\author{
  J Dudek\email{Jerzy.Dudek@unistra.fr},
  D Curien\email{Dominique.Curien@iphc.cnrs.fr} and
  D Rouvel\email{David.Rouvel@iphc.cnrs.fr} \\
  \it Institut Pluridisciplinaire Hubert Curien, IN2P3-CNRS, France,
  and Universit\'e de Strasbourg,\\ 
  \it 23, rue du Loess, B. P. 28, F-67037 Strasbourg Cedex 2, France\\
  K Mazurek\email{Katarzyna.Mazurek@ifj.edu.pl} \\
  \it The Niewodnicza\'nski Institute of Nuclear Physics,
  Polish Academy of Sciences\\
  \it ul. Radzikowskiego 152, Pl-31432 Krak\'ow, Poland\\
  Y R Shimizu\email{Shimizu@phys.kyushu-u.ac.jp}~ and 
  S Tagami\email{Tagami@phys.kyushu-u.ac.jp}\\
  \it Department of Physics, Faculty of Sciences,
  \it Kyushu University, Fukuoka 812-8581, Japan
}

\pacs{21.10.-k, 21.10.Pc, 21.30.-x, 21.60.-n, 21.60.Ka}

\date{}

\maketitle

\begin{abstract}

We discuss the predictions of the large scale calculations using the realistic realisation of the phenomenological nuclear mean-field theory. Calculations indicate that certain Zirconium nuclei are tetrahedral-symmetric in their ground-states. After a short overview of the research of the nuclear tetrahedral symmetry in the past we analyse the predictive capacities of the method and focus on the $^{96}$Zr nucleus expected to be tetrahedral in its ground-state. 

\end{abstract}
  
%%%%%%%%%%%%%%%%%%%%%%%%%%%%%%%%%%%%%%%%%%%%%%%%%%%%%%%%%%%%%%%%%%%%%%%%%%%%%%%%

\section{Nuclear Point-Group Symmetries: The Search of Tetrahedral Symmetry -- A Short Overview}
\label{Sect.01}

Analysing point-group symmetries of molecules has become one of the standard tools of molecular quantum mechanics. These symmetries often result from the relative positions of the constituent atoms (as e.g.~positions of the Hydrogen atoms in the CH$_4$ molecule) and in this sense their presence may be considered rather intuitive. In contrast, there seem to be no intuitively direct analogies with atomic nuclei, which can be considered compact objects whose volume is nearly equal to the sum of the volumes of the constituent nucleons. Moreover, the underlying strong interactions, both non-central and non-local, are much more complex compared to the Coulomb interactions governing the molecular structure. 

Despite that, the molecular-geometry guided intuition has been followed by certain authors, who constructed group-theoretical models of nuclei based on the idea of the modelling in terms of alpha clusters. As an example, in analogy to the tetrahedral symmetry of the methane molecules, the nuclear tetrahedral symmetry induced by four $\alpha$-clusters in $^{16}$O has been discussed in e.g.~Ref.\,\cite{Rob82}. 

Nuclear mean-field theory is one of the most successful tools in nuclear structure. Today, the most frequent realisations of this theory are:  a.~The Phenomenological one (the so-called macroscopic-microscopic method); b.~The  Relativistic Mean Field theory based on the Dirac formalism, and: c.~The Hartree-Fock theory. The first of them is technically the best adapted to study the nuclear geometrical symmetries. This is in particular true for the phenomenological Woods-Saxon realisation of the approach, according to which the central potential is constructed as
\begin{equation}
   V_{WS}(\vec{r};V_0,R_0,a_0)
   \equiv
   \frac{V_0}{1+\exp[{\rm dist}_\Sigma(\vec{r};R_0)/a_0]},
                                                                 \label{eqn.01}
\end{equation}
where $\Sigma$ denotes the nuclear surface and ${\rm dist}_\Sigma(\vec{r};R_0)$ the distance of the point $\vec{r}$ from the surface. Above, $V_0$, $R_0$ and $a_0$ are adjustable parameters characterising the potential depth, radius and diffusivity, respectively. In our approach, they are fixed once for all, i.e., for all the nuclei in the Periodic Table and independent of the deformation  -- where from the name: `Universal Parameterisation'. 

According to the above definition, the potential contains a constant diffusivity parameter and therefore has an overall linear dependence of the argument of the exponential on the distance of a given point $\vec{r}$ from the nuclear surface. [This is in contrast to some alternative forms used in the literature, in which the diffusivity is treated as a position-dependent quantity.] It follows that the resulting Hamiltonian has exactly the same symmetry as the underlying nuclear surface. Of course similar can be said about the spin-orbit potential of the Woods-Saxon type. Consequently, the analysis of the point-group symmetries of the Woods-Saxon mean-field Hamiltonian can be reduced to the analysis of the point-group symmetries of the underlying nuclear surfaces. 

With the help of such an approach it was shown for the first time in 
Ref.\,\cite{XLi94}, that the non-trivial nuclear point-group symmetries, such as the tetrahedral one, can be easily realised with the help of the realistic nuclear Woods-Saxon Hamiltonians with the single deformation parameter $\alpha_{32}$. The calculated single-particle nucleonic spectra were shown to satisfy the two-fold and four-fold degeneracies related to the $E$, and $E^*$ (two-dimensional) and $G$ (four-dimensional) irreducible representations of the tetrahedral group, cf.~figures 3 and 4 in the above reference. Moreover, it was shown that the local minima on the total potential energy surfaces corresponding to the tetrahedral symmetry are obtained thus paving the way towards the idea of the new, richer forms of the shape coexistence in atomic nuclei. Such a coexistence may involve non-axial, i.e.~different from the prolate/oblate quadrupole or pear-shape octupole symmetries discussed so far abundantly in the literature. 

At the same time it has been suggested \cite{XLi94} that the new point-group symmetries generate the new chains of magic numbers in analogy to the well known ones, $(Z/N)_{\rm spherical}=$8, 20, 28, 50, 82 and 126 generated by the spherical symmetry. The prediction of the tetrahedral magic numbers $(Z/N)_{\rm tetrahedral}=$56, 64, 70, 90, 112 and 136 has been formulated in the cited article where only moderately heavy and heavy nuclei have been studied.

Extending realistic calculations which involve tetrahedral symmetry of nuclear shapes represented by the $T_d$-group and $\alpha_{32}$ deformation parameter, from now on referred to as tetrahedral, it was shown that a combination of quadrupole and tetrahedral components may lead to a new class of shapes, symmetric with respect to the $D_{2d}$ group. The latter describes, in the nuclear context, some exotic superdeformed shapes, with the spherical harmonic expansion in terms of the leading axial-quadrupole, $\alpha_{20}$, and tetrahedral, $\alpha_{32}$, components combined, as suggested in Ref.\,\cite{XLi91} [cf.~also Ref.\,\cite{JSk92}, the latter focussed on the Hg-Pb region, and Ref.\,\cite{JSk91}.] 

Unfortunately, the names used sometimes in the literature to describe the related nuclear geometry such as `non-axial octupole shapes', hide the presence of possibly very distinct symmetry effects\footnote{Let us recall that atomic nuclei (as all other quantum systems) undergo what is referred to as `zero-point motion', perpetual oscillations, among others, in all the shape degrees of freedom, what implies that the {\em static} symmetries described by the mean-field Hamiltonian cannot represent more than just the leading order symmetries.}. Indeed, the nuclear octupole deformations $\alpha_{31}$, $\alpha_{32}$ and $\alpha_{33}$ imply very different point-group symmetries. It then may become useful to design distinct spectroscopic criteria associated with each of these symmetries, e.g.~in terms of the energy-vs.-spin staggering (see below), approximate level degeneracies or specific branching ratios among the electromagnetic transitions -- all these depending on the dominating$^1$ symmetry. 

The mean-field theory studies were continued in Ref.\,\cite{JDu02}, where the tetrahedral-symmetry induced `magic' gaps in the single-particle nuclear spectra have been found to include $(Z/N)_{\rm tetrahedral}=$16, 20, 32, and 40 for the relatively light nuclei, and 142 for the heaviest ones. In this context the role of the four-fold degeneracies, alternatively, the four-dimensional irreducible representations of the tetrahedral group, as the background of the large tetrahedral shell gaps has been emphasised. Moreover, the predictions related to the new forms of (tetrahedral) shape-isomerism have been formulated for $^{80}$Zr, $^{108}$Zr, $^{160}$Yb and $^{242}$Fm. 

These calculations have been extended in Ref.\,\cite{NSc04}, employing the Skyrme Hartree-Fock technique with the SLy4 parameterisation. These results suggested, that the very exotic $^{110-112}$Zr nuclei are tetrahedral in their ground-states [cf.~also Ref.\,\cite{NSc4a}]. In Ref.\,\cite{NSc04}, yet another research direction has been proposed and illustrated, {\em viz.}~the application of the group representation theory to calculate the symmetry-induced characteristic intensity branching ratios of the electromagnetic transitions emitted by the tetrahedral-symmetric nuclear quantum rotor [cf.~Fig.\,(4) in Ref.\,\cite{NSc04}]. 

A more general `theory of nuclear stability' based on the point-group symmetries has been formulated in Ref.\,\cite{JDu08}, where, moreover, the illustrations related to the tetrahedral symmetry minima in $^{80}$Zr, $^{96}$Zr and $^{110}$Zr nuclei can be found, together with the indication that the axial-symmetry octupole-shape minima are in a direct competition with the tetrahedral ones -- all at the zero quadrupole deformation. These calculations indicate that the nuclei in question should manifest strong octupole transitions and thus strong $B({\rm E3})$ values {\em which according to theory, certainly do not correspond to the quadrupole-(super)deformed minima predicted in $^{80}$Zr and $^{110}$Zr}.

Independently, the Hartree-Fock mean-field calculations focused on the $Z=N$ nuclei in Ref.\,\cite{MYa01}, confirmed both the presence of the superdeformation and the instability of the spherical configuration with respect to the 
$\alpha_{32}$ deformation in the tetrahedral doubly-magic $^{80}_{40}$Zr nucleus. The tetrahedral instability of the spherical configurations in selected nuclei, including Zirconium has been confirmed in \cite{POl06} where, in addition, the sensitivity of the predictions with respect to the choice of the Skyrme Hartree-Fock parameterisation (SIII, SkP, SkM$^*$) has been tested. Calculations using again Skyrme Hartree-Fock method supplemented with the Generator Coordinate Ansatz for $^{80}$Zr and $^{98}$Zr \cite{KZb06} confirmed the competition between the deformations of the octupole-axial and tetrahedral symmetries, the tetrahedral one dominating. 
\begin{table}[h]
\caption[TT]{Experimental values of the $B({\rm E3})$ reduced transition probabilities
             in Weisskopf units for to the first $3^-$ excitation, 
             in the vicinity of the tetrahedral doubly-magic nucleus
             $^{96}_{40}$Zr$^{}_{56}$ showing the strongest effect in $^{96}$Zr. 
             The data are from Ref.\,\cite{EData}.
                                                                \label{tab.01}}
\begin{center}
\begin{tabular}{||l||c|c|c|c||} \hline\hline
\phantom{\bigg|}  Z\,vs.\,N    & 54 & 56  & 58 & 60 \\
                                                      \hline\hline
$_{\;\;46}^{}$Pd  & - & \phantom{\bigg|}- & - &29$\pm$10 \\
$_{\;\;44}^{}$Ru  & - & 14$\pm$3 & - & -        \\[1mm]
$_{\;\;42}^{}$Mo  & - & 31$\pm$4 & 35$\pm$3 & -        \\[1mm]
$_{\;\;40}^{}$Zr  & - & {\color{red}57$\pm$4} & - & -        \\[1mm]
$_{\;\;38}^{}$Sr  & 18.3$\pm 11$ & - & - & -        \\[1mm]
                                                      \hline\hline
\end{tabular} 
\end{center}
\end{table}

Among the three doubly-magic tetrahedral nuclei, i.e., $^{80}_{40}$Zr$_{40}$, 
$^{96}_{40}$Zr$_{56}$ and $^{110}_{\;\;40}$Zr$_{70}$, only the second one is stable. As the consequence, the experimental results concerning the reduced probabilities for the $B({\rm E3})$ transitions can be found in the literature only for this one, and for a few neighbouring nuclei.  
The existing results are presented in Table \ref{tab.01}, showing that the measured value for $^{96}$Zr, i.e., $B({\rm E3})=$57$\pm$4 W.u., clearly dominates. To our knowledge this is the largest $B({\rm E3})$ value ever measured. Given the fact that the theory predictions favour the dominance of the $\alpha_{32}$ over $\alpha_{30}$, cf.~also the results in Figs.\,\ref{fig.03}-\ref{fig.04} below, one is tempted to suggest that these very big values should be attributed to the presence of the tetrahedral rather than octupole-axial symmetry. [For the domination of the $\alpha_{32}$ over $\alpha_{30}$ deformation, the reader is referred to Fig.\,(4) in Ref.\,\cite{JDu08} and to Ref.\,\cite{KZb09}.] Finally, an overview of the tetrahedral symmetry oriented nuclear shell-effect calculations focussing on the Rare Earth nuclei can be found in Ref.\,\cite{JDu06}.

More recent microscopic calculations addressing the issue of the nuclear tetrahedral symmetry have been performed using advanced projection techniques \cite{STa12} including the Generator Coordinate approach and various forms of the nuclear interactions varying between the phenomenological and the self-consistent Gogny Hartree-Fock approach. These calculations fully confirm the importance of the tetrahedral symmetry on the nuclear level, the symmetry which strongly enhances the nuclear stability properties. The calculations demonstrate in particular the existence of the privileged spin-parity combinations of the nuclear states forming the tetrahedral sequences (`bands'), 
the structures whose properties will be helpful in analysing the results of dedicated experiments, cf.\,Ref.\,\cite{STa13} and more particularly Ref.\,\cite{ST13a}. 

%%%%%%%%%%%%%%%%%%%%%%%%%%%%%%%%%%%%%%%%%%%%%%%%%%%%%%%%%%%%%%%%%%%%%%%%%%%%%%%%
%%%%%%%%%%%%%%%%%%%%%%%%%%%%%%%%%%%%%%%%%%%%%%%%%%%%%%%%%%%%%%%%%%%%%%%%%%%%%%%%

\section{Testing the Method's Prediction Capacity in Terms of the Nuclear 
         Non-axial Configurations }
\label{Sect.02}

Whereas the nuclear quadrupole deformations are characterised by a single non-axiality parameter, $\alpha_{22}$, the octupole non-axial degrees freedom are characterised by three: $\alpha_{31}$, $\alpha_{32}$ (tetrahedral) and 
$\alpha_{33}$. According to the shape parameterisation in terms of the spherical harmonics,  
\begin{figure}[ht!]
   \includegraphics[width=1.1\columnwidth]{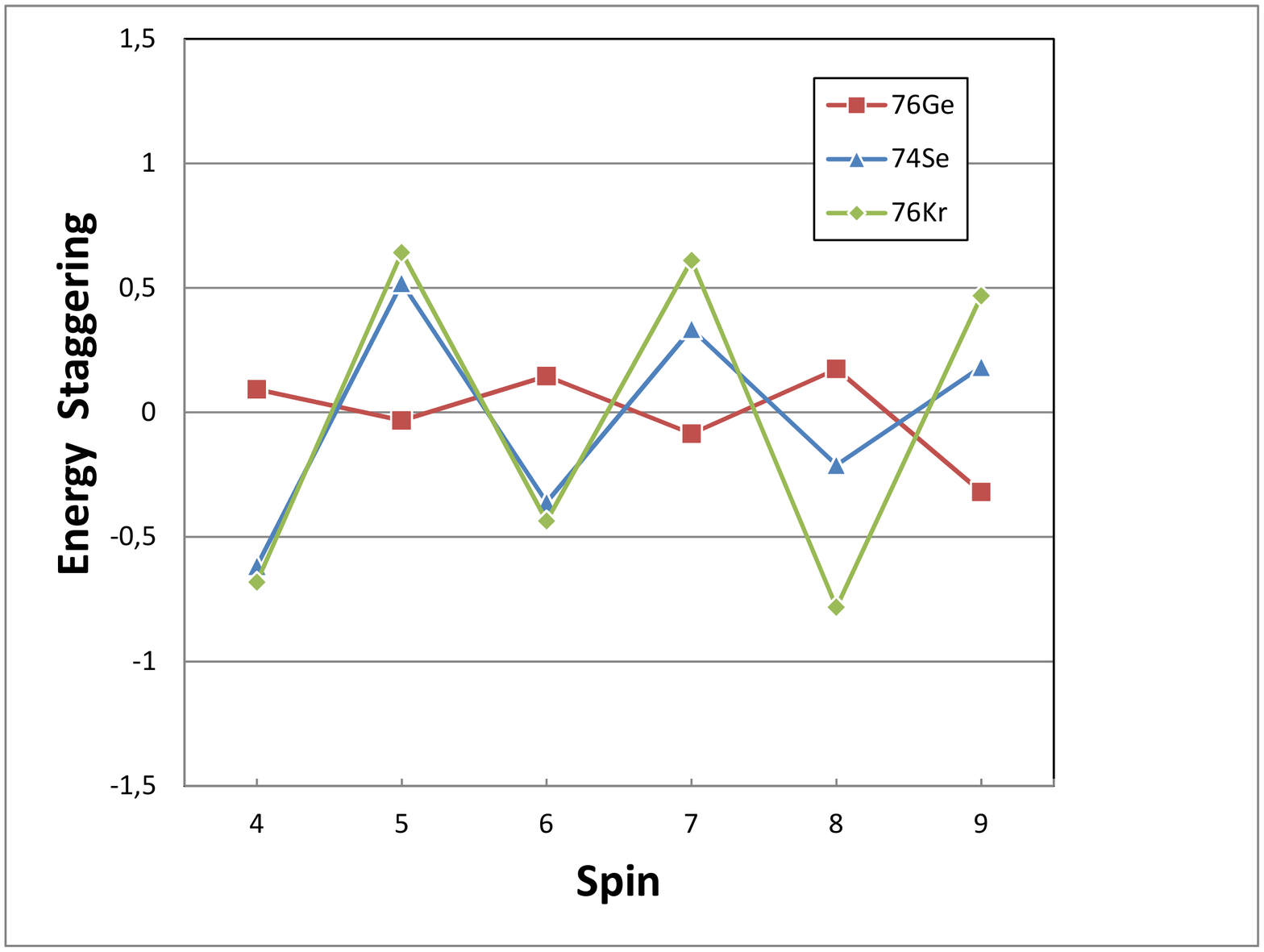}\vspace{-5mm}
   \caption{Staggering properties of the nuclei surrounding 
            the recently measured $^{76}$Ge, Ref.\,\cite{Tot13}. The data for
            $^{74}$Se and $^{76}$Kr are from \cite{D74Se} and \cite{D76Kr},
            respectively.} 
                                                               \label{fig.01}
\end{figure}
\begin{figure}[ht!]
   \includegraphics[width=\columnwidth]{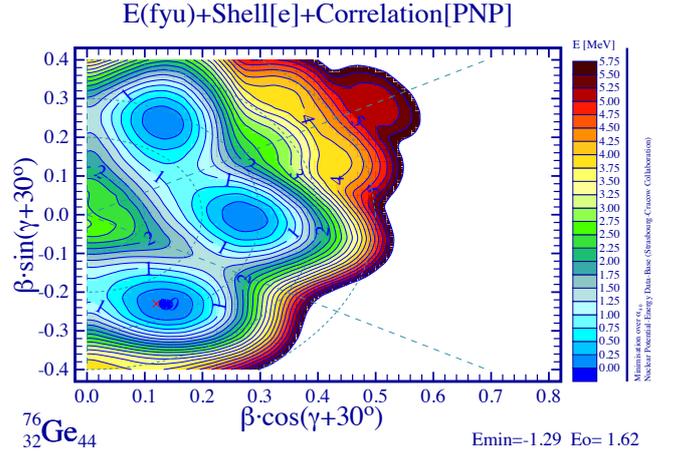}\vspace{-5mm}

   \includegraphics[width=\columnwidth]{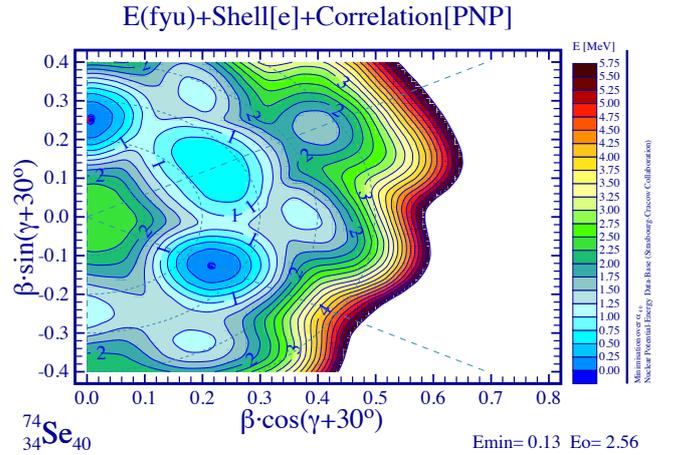}\vspace{-5mm}

   \includegraphics[width=\columnwidth]{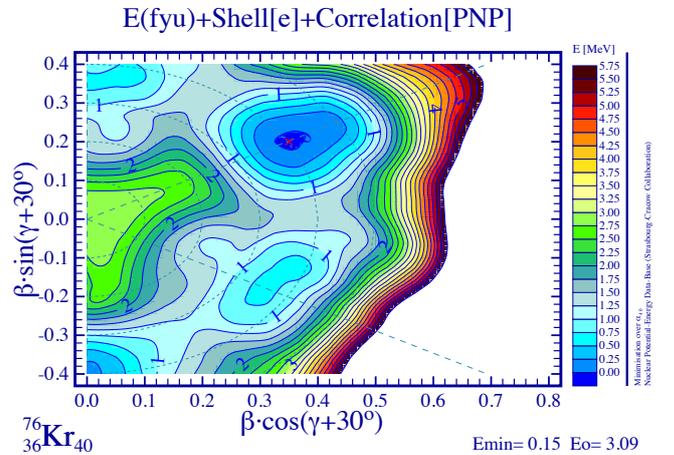}
   \caption{Total energy surfaces of the nuclei illustrated in 
            Fig.\,\ref{fig.01}. 
            At each $(\beta,\gamma)$-point the energy was minimised over the 
            hexadecapole $\alpha_{40}$ deformation, where from the `traditional' 
            symmetry in terms of sectors of $\Delta \gamma = 60^o$ is lost. Top:
            $^{76}$Ge, whose stable triaxial deformation has recently
            been confirmed experimentally, cf.~Ref.\,\cite{Tot13}, and 
            Fig.\,\ref{fig.01}. According to our calculations the equilibrium 
            value $\gamma_{eq.}\approx 30^o$. For $^{74}$Se, middle, and
            $^{76}$Kr, bottom, there exist only the axial symmetry minima which
            may give rise to the rotational bands observed.} 
                                                                  \label{fig.02} 
\end{figure}
the non-axiality is determined by the $\varphi$-dependence of the  spherical harmonics
$
  Y_{\lambda\mu}(\vartheta,\varphi)
  \propto 
  P_\lambda^\mu(\cos\vartheta)\exp(i\mu\varphi)
$
and it follows that this dependence is identical for the non-axial quadrupole and tetrahedral symmetry shapes.

Since no single case of the tetrahedral symmetry has been demonstrated so far through experiment we will illustrate, following Refs.\,\cite{NVZ91,Tot13}, the prediction capacity of the present method using the results for the quadrupole-type non-axiality, obtained prior to the recent experiment of Ref.\,\cite{Tot13}. This is pertinent in the actual context since we will be able to compare the detailed shape-coexistence properties in the nuclei which are close to the Zirconium nuclei addressed in this article. 

We follow here the criterion based on the early Ref.\,\cite{NVZ91}, whose authors introduced the energy vs.~spin staggering properties to distinguish between the rigid triaxial quantum rotors [as discussed long ago by Davydov and Filippov, \cite{ASD58}] and the axially symmetric rotor with possibly triaxial vibrations (`$\gamma$-soft rotor') \cite{WLJ56}. According to those models, the odd-spin (odd-$I$) states of a $\gamma$-soft rotor-band are closer in energy to the neighbouring even-spin states with $(I+1)$ rather than $(I-1)$; the opposite tendency is true for the triaxial rotor. To exhibit these tendencies numerically one introduces the double difference
\begin{equation}
   \mathcal{S}(I)\equiv \textstyle \frac{1}{2}
   \left\{
         [E(I+1) - E(I)] -[E(I) - E(I-1)]
   \right\}
\end{equation}
which represents the finite difference approximation to the second derivative of the energy vs.~spin, $\frac{\partial^2 E}{\partial I^2}$, thus displaying the sign differences predicted by the rotor models mentioned. To illustrate the experimental results in Fig.\,(\ref{fig.01}) we have renormalised the above definition to comply with the convention used in \cite{Tot13} by using
$S(I) \equiv 2\mathcal{S}(I-1) / E(2^+_1)$. 

The results presented in the Figure demonstrate that according to the model-criterion just specified, the nucleus $^{76}$Ge exhibits the properties of the rigid rotor whereas the close-neighbours, for which the relatively unambiguous experimental data exist are expected to be axial or nearly axial in agreement with the theoretical predictions illustrated in Fig.\,\ref{fig.02}. We may conclude that performance of the model calculations can be qualified as realistic; we believe that so are the predictions for the nucleus $^{96}$Zr which will be discussed next. 

%%%%%%%%%%%%%%%%%%%%%%%%%%%%%%%%%%%%%%%%%%%%%%%%%%%%%%%%%%%%%%%%%%%%%%%%%%%%%%%%

\section{Competing Two Octupole Modes: \boldmath $^{96}$Zr Case}
\label{Sect.03}

The calculations by the present-, and by the authors cited above leave no doubt that: a.~The two octupole modes, i.e., $\alpha_{32}$ representing the tetrahedral ($T_d$) symmetry, and $\alpha_{30}$ representing axial ($C_\infty$) symmetry compete on the total energy maps in the Zirconium region, and b.~The tetrahedral symmetry most often wins the competition. The strong presence of the collective octupole mode in the region as manifested by the experiment, 
cf.~Table \ref{tab.01} and surrounding
\begin{figure}[ht!]
   \includegraphics[width=\columnwidth]{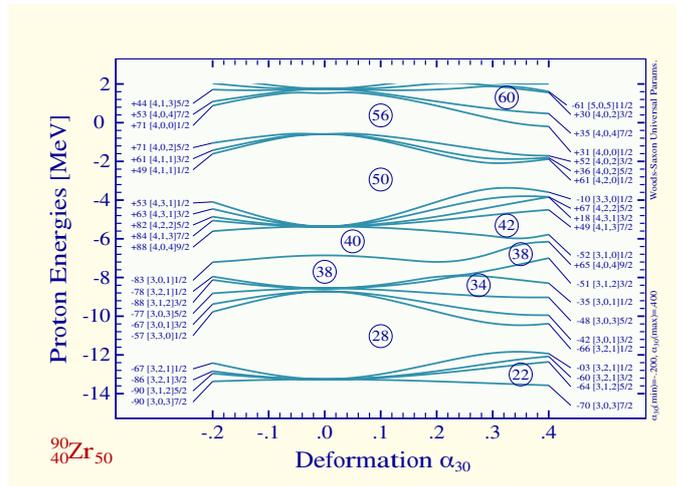}
   \caption{Proton single-particle levels as a function of the 
            axial-symmetry octupole deformation, $\alpha_{30}$, calculated using
            the deformed Woods-Saxon Hamiltonian with the `universal' parameter
            set. The Nilsson labels are supplemented with the symbol $\kappa$,
            [i.e. they have the form $\kappa[N n_z \Lambda]\Omega$]
            representing the square of the amplitude of probability 
            of the corresponding basis state in the WS wave function, whereas 
            the sign of $\kappa$ refers to the expected value of the parity 
            operator in each WS solution. For the sake of this illustration all
            other deformation parameters are fixed at 0. 
            } 
                                                                \label{fig.03} 
\end{figure}
\begin{figure}[ht!]
   \includegraphics[width=\columnwidth]{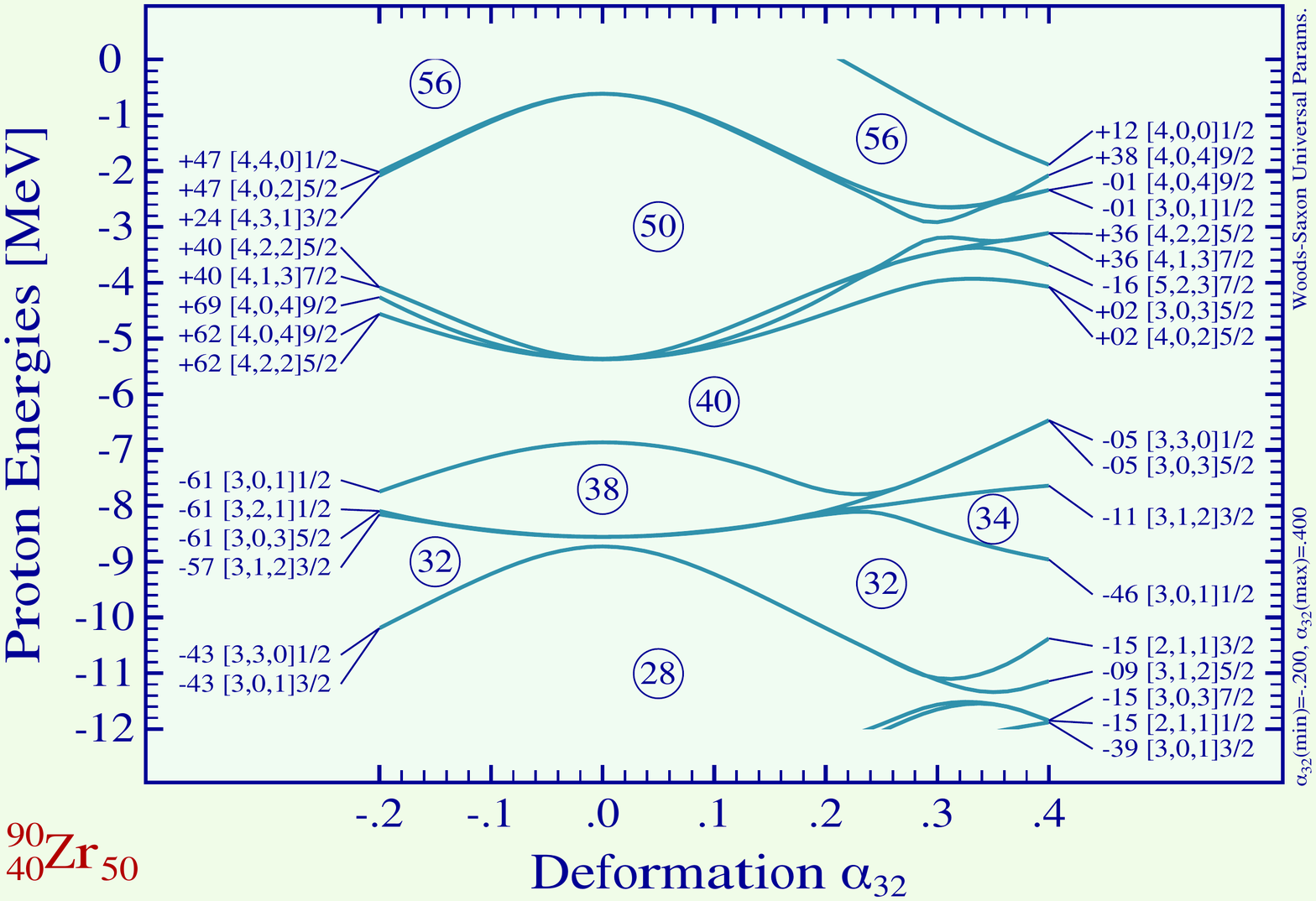}
   \caption{Similar to the preceding one but for the tetrahedral deformation
            $\alpha_{32}$; both diagrams are symmetric with respect to the 
            zero argument value. The argument displayed within the interval 
            $[-0.2,+0.4]$ allows to judge the degree of parity-mixing with 
            increasing octupole deformations by comparing the $\kappa$-values
            at the argument values 0.2 and 0.4\,. [For more details 
            cf.~Ref.\,\cite{NSPhD}.]} 
                                                                \label{fig.04} 
\end{figure}
text, clearly confirms this prediction. It then follows that the tetrahedral symmetry should be the dominating factor, and it will be instructive to illustrate the microscopic origin of this mechanism.

Figures \ref{fig.03} and \ref{fig.04} illustrate the characteristic dependence of the proton single-particle Woods-Saxon energies [the neutron levels present very similar features and are not shown here] on the two competing deformations. Observe that whereas at tetrahedral symmetric shapes the gap at $Z=40$ extends to the very large deformations of about $\alpha_{32}\approx 0.3$, the same gap decreases continuously as a function of $\alpha_{30}$. Similar observations can be made about the gap at $Z=56$ (in the case of protons this gap is close to the zero-binding limit what is not the case for the neutrons). The origin of this systematic difference can be traced back to the four-fold degeneracies of the majority of the single-particle levels in the case of the tetrahedral symmetry.
The corresponding levels can be identified in Fig.\,\ref{fig.04} as marked with the double Nilsson labels. Because of the fact that the dominating majority of levels belong to this category the average inter-level spacing is systematically larger in the tetrahedral symmetry case leading to overall larger spacings.
\begin{figure}[ht!]
\begin{center}
   \includegraphics[width=0.7\columnwidth]{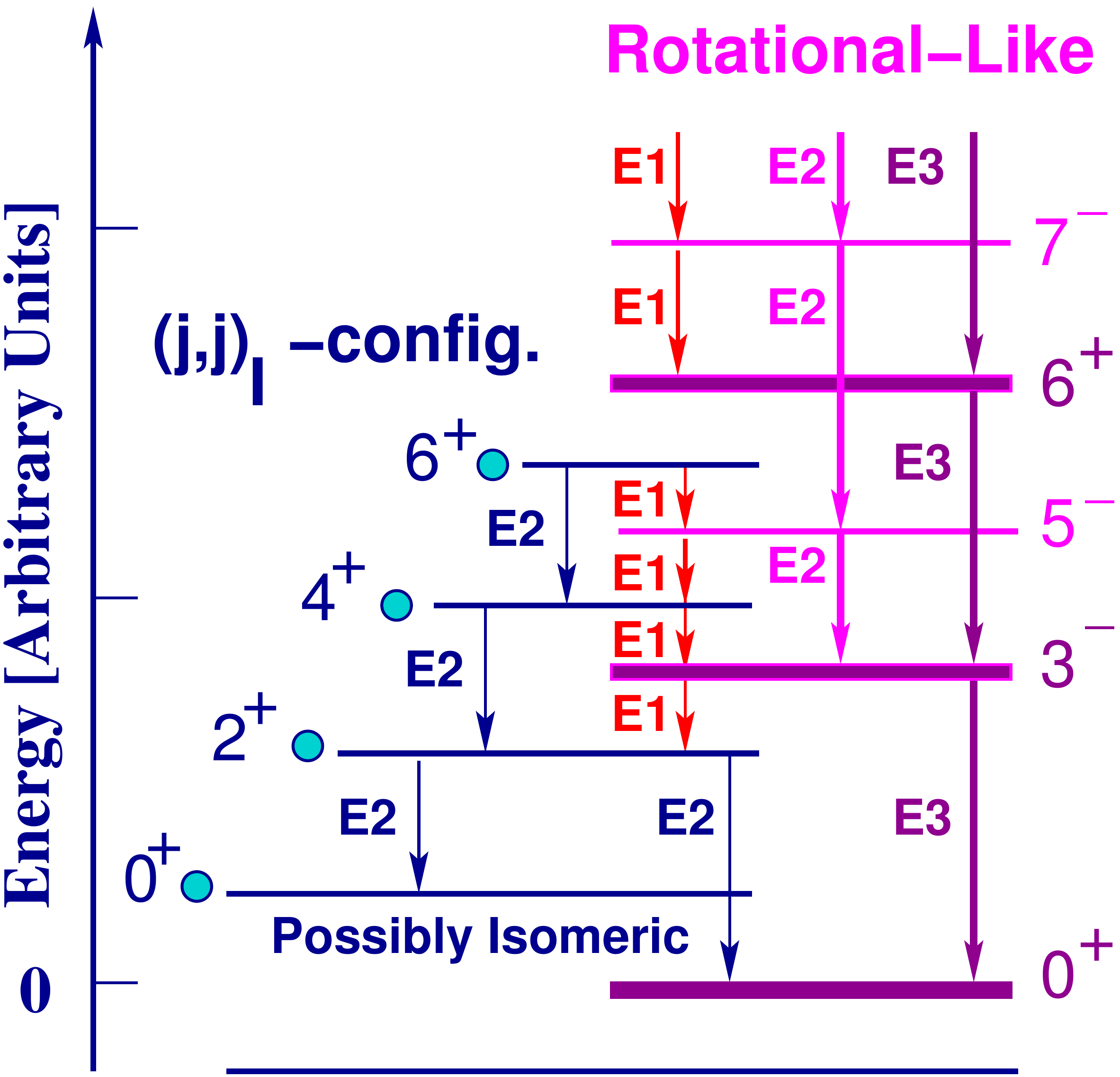}
   \caption{Schematic. The lowest energy-and-spin part of the excitation scheme
            expected on the basis of our calculations. At the deformation of 
            $\alpha_{32}\approx 0.15$, the collective rotational transition
            energies are expected to be approximately linear in spin 
            \cite{STa13,ST13a} rather than proportional to $I(I+1)$ - where from
            the term `rotational-like'. The E3-transitions can be excited from the tetrahedral ground-state via Coulomb excitations.} 
                                                                \label{fig.05} 
\end{center}
\end{figure}

According to our calculations, the ground-state minimum of $^{96}$Zr corresponds to the tetrahedral deformation with $\alpha_{32}\approx0.15$ and a small octahedral deformation, with the energy about 1300 keV below the spherical minimum\footnote{Our total energy as a function of tetrahedral deformation, is minimised  over the octahedral deformation since both symmetries have, from the mathematics point of view, strong similarities. This may be one of the reasons that our energy minima are better pronounced and the energy estimates are lower as compared to some other approaches which ignore this mechanism.}. At the strict symmetry limit the excitation (and feeding) of the ground-state through collective transitions coming from either rotational or vibrational excited states is possible {\em only via the octupole $E3$ transitions} since the tetrahedral deformed quantum objects have vanishing collective quadrupole and dipole moments. However, such a state may receive transitions of the  single-particle strengths from the non-collective particle-hole excitations. The lowest-lying non-collective excitations are expected to come either from the tetrahedral ground-state or from the excited spherical energy minimum giving rise to the $E(I=\{j^2\}_I)$ sequences as indicated schematically in Fig.\,\ref{fig.05}. Since the highest-$j$ neutron orbital above the Fermi level is $g_{7/2}$ one may expect such a sequences to terminate at $I^\pi=6^+$ states. The dedicated analysis of the experimental data in this nucleus including the reduced transition probabilities, negative and positive parity rotational-like sequences and their possible interplay, partly following the lines discussed in \cite{ST13a}, is in progress and will be published elsewhere.

%%%%%%%%%%%%%%%%%%%%%%%%%%%%%%%%%%%%%%%%%%%%%%%%%%%%%%%%%%%%%%%%%%%%%%%%%%%%%%%%
%%%%%%%%%%%%%%%%%%%%%%%%%%%%%%%%%%%%%%%%%%%%%%%%%%%%%%%%%%%%%%%%%%%%%%%%%%%%%%%%

\section{Possible Impact of Tetrahedral Symmetry}
\label{Sect.04} 

Because of the unusual four-fold degeneracies of single-nucleon levels, tetrahedral symmetry of the mean-field is expected to generate strong shell effects and thus relatively strongly bound ground- or shape-isomeric states. This is expected to justify the presence of the new waiting-point nuclei which would help to explain missing elements of the nucleosynthesis models and modify the actually known nuclear abundance scheme -- in addition to paving the way to new spectroscopic features, new ideas about the structure (and the very definition) of the rotational bands [cf.~Ref.\,\cite{ST13a} in these proceedings] with the new selection rules for the collective rotational transitions. 

%%%%%%%%%%%%%%%%%%%%%%%%%%%%%%%%%%%%%%%%%%%%%%%%%%%%%%%%%%%%%%%%%%%%%%%%%%%%%%%%
%%%%%%%%%%%%%%%%%%%%%%%%%%%%%%%%%%%%%%%%%%%%%%%%%%%%%%%%%%%%%%%%%%%%%%%%%%%%%%%%

\begin{ack}
This work is supported by the LEA COPIGAL project 04-113 and by the COPIN-IN2P3 agreement No.06-126.
\end{ack}

%%%%%%%%%%%%%%%%%%%%%%%%%%%%%%%%%%%%%%%%%%%%%%%%%%%%%%%%%%%%%%%%%%%%%%%%%%%%%%%%


\begin{thebibliography}{99}

\bibitem{Rob82} Robson D 1982
                {\it Phys.~Rev.}~{\bf C 25} 1108
                
\bibitem{XLi94} Li X and Dudek J 1994 
                {\it Phys.~Rev.}~{\bf C 49}, R1250
                          
\bibitem{XLi91} Li X, Dudek J and Romain P 1991\\
                {\it Phys.~Lett.}~{\bf B 271} 281

\bibitem{JSk92} Skalski J 1992
                {\it Phys.~Lett.}~ {\bf B 274} 1

\bibitem{JSk91} Skalski J 1991
                {\it Phys.~Rev.}~{\bf C 43}, 140 

\bibitem{JDu02} Dudek J, G\'o\'zd\'z A, Schunck N and Mi\,skiewicz M 2002
                {\it Phys.~Rev.~Lett.}~{\bf 88}, 252502 

\bibitem{NSc04} Schunck N, Dudek J, G\'o\'zd\'z A and Regan P H 2004
                {\it Phys.~Rev.}~{\bf C 69}, 061305(R)

\bibitem{NSc4a} Schunck N and Dudek J 2004\\
                {\it Int.~J.~Mod.~Phys.}~{\bf E 13} 213

\bibitem{JDu08} Dudek J, Mazurek K, Curien D, Dobrowolski A, G\'o\'zd\'z A, 
                Hartley D, Maj A, Riedinger L and Schunck N 2008
                {\it Acta Physica Polonica} {\bf B 40} 713

\bibitem{MYa01} Yamagami M, Matsuyanagi K, and Matsuo M 2001
                {\it Nucl.~Phys.}~{\bf A 693} 579

%               10 references above

\bibitem{POl06} Olbratowski P, Dobaczewski J, Powalowski P, Sadziak M 
                and Zberecki K 2006\\
                {\it Int.~J.~Mod.~Phys.}~{\bf E 15} 333

\bibitem{KZb06} Zberecki K, Magierski P, Heenen P H and Schunck N 2006
                {\it Phys.~Rev.}~{\bf C 74} 5 

\bibitem{EData} National Nuclear Data Center @ www.nndc.bnl.gov


\bibitem{KZb09} Zberecki K, Heenen P H and Magierski P 2009
                {\it Phys.~Rev.}~{\bf C 79} 014319

\bibitem{JDu06} Dudek J, Curien D, Dubray N, Dobaczewski J, Pangon~V,
                Olbratowski~P and Schunck~N 2006
                {\it Phys.~Rev.~Lett.}~{\bf 97} 072501

\bibitem{STa12} Tagami~S and Shimizu~Y~R 2012\\
                {\it Prog.~Theor.~Phys.} {\bf 127}, 79

\bibitem{STa13} Tagami~S, Shimizu~Y~R and Dudek~J 2013\\
                {\it Phys.~Rev.}~{\bf C\,87} 054306 

\bibitem{ST13a} Tagami~S, Shimada~M, Fujioka~Y, Shimizu~Y~R and Dudek J 2013, 
                {\it These Proceedings}

\bibitem{NVZ91} Zamfir N V and Casten R F, 1991,\\
                {\it Phys.~Lett.}~{\bf B 260} 265

\bibitem{Tot13} Toh Y, {\em et al.}, 2013,
                {\it Phys.~Rev.}~{\bf C 87} 041304
      
%               20 references above

\bibitem{ASD58} Davydov A S and Filippov S, 1958,
                {\it Nucl.~Phys.}~{\bf 8} 237
      
\bibitem{WLJ56} Wilets L and Jean M, 1956,
                {\it Phys.~Rev.}~{\bf 102} 788

\bibitem{D74Se} Singh B and Farhan A R (2006) 
                {\it Nucl.~Data Sheets} {\bf 107} 1923;  
                D\"oring J {\em et al.,} 1998
                {\it Phys.~Rev.}~{\bf C 57} 2919
   
\bibitem{D76Kr} Valiente-Dob\'on J J{\em et al.,} 2005\\
                {\it  Phys.~Rev.}~{\bf C 71} 034311      

\bibitem{NSPhD} Schunck N, Ph.-D. thesis, Strasbourg University,\\
                http://tel.archives-ouvertes.fr/?langue=en
                
\end{thebibliography}
\end{document}